\documentclass[twocolumn,amsmath,amssymb,floatfix,aps,prl]{revtex4}  
  
\usepackage{graphicx}
\usepackage{epsfig}   
\usepackage{dcolumn}
\usepackage{bm}
\usepackage{times}  
  
\begin{document}   
  
  
\title{Free Magnetic Moments in Disordered Metals}   
  
\author{Stefan Kettemann$^{1,2}$}   
  
\author{Eduardo R. Mucciolo$^{3}$}   
  
\affiliation{$^1$ Institut f\"ur Theoretische Physik, Universit\" at  
Hamburg, Jungiusstra\ss e 9, 20355 Hamburg, Germany, }  
  
\affiliation{$^2$ Max-Planck Institute for Physics of Complex Systems,  
 N\" othnitzer Stra\ss e 38, Dresden, Germany,}  
   
\affiliation{$^3$ Department of Physics, University of Central  
Florida, Box 162385, Orlando, FL 32816-2385, USA,}

\date{November 23, 2005}  
  
\begin{abstract}  
The screening of magnetic   moments in metals, the Kondo effect, is
found to be quenched with a finite probability in the presence of
nonmagnetic disorder. Numerical results for a disordered electron
 system show that the distribution of Kondo temperatures deviates
 strongly from the result expected from random matrix theory. A
 pronounced second peak emerges for small Kondo temperatures, showing
 that the probability that magnetic moments remain unscreened at low
temperatures increases with disorder. Analytical calculations, taking
into account correlations between eigenfunction intensities yield a
finite width for the distribution in the thermodynamic
limit. Experimental consequences for disordered mesoscopic metals are
discussed.

\end{abstract}  
  
  
\maketitle


 In a metal with antiferromagnetic exchange interaction between a local
magnetic moment and the conduction electrons, correlations cause a
change in the Fermi liquid ground state. The screening of the
localized spin by the formation of a Kondo singlet below the Kondo
temperature, $T_K$, is affected by disorder in various
ways. Fluctuations in the exchange coupling due to random positioning
of magnetic moments results in a dispersion of $T_K$
\cite{bhatt}. Since $T_K$ is defined by an integral equation similar
to the BCS equation for the critical temperature of a superconductor,
one could expect, by analogy,  the Anderson theorem \cite{rickayzen}
to be valid. In that case, the leading correction to $T_K$ would be of
order $T_K^{(0)}/g$, where $T_K^{(0)}$ is the bulk Kondo temperature
in the clean limit and $g = E_F \tau$, the dimensionless conductance,
is the ratio between the Fermi energy $E_F$ and the elastic scattering
rate $1/\tau$. However, the equation for $T_K$ requires a sum over the
local density of states (LDOS) rather than the global one (DOS).
It is known that disorder induces correlations in energy in the LDOS
\cite{mirlin}. Thus, impurities
cause additional fluctuations on the LDOS and thereby on $T_K$. 


Let us begin by describing the relevant approaches to the different
regimes in the problem. Consider a single magnetic impurity in a
closed, phase-coherent metallic grain where the energy levels are
discrete, a Kondo box \cite{kroha,meraikh}, with mean level spacing
$\Delta = 1/\nu\, L^d$ and Thouless energy $E_c=D_e/L^2$. Here, $L$ is
the system linear size, $\nu$ the DOS, $D_e$ the diffusion constant,
and $d$ the space dimension. We note that $T_K$ does not exceed
$1/\tau$ in metals, ruling out perturbation theory in
$1/\tau$. However, the regime $E_c < T_K < 1/\tau$ applies to a wide
range of metals and an expansion in terms of diffusion diagrams is
permitted in that case \cite{shklovskii}. For smaller grains, when
$\Delta < T_K < E_c$, random matrix theory (RMT) yields a distribution
of $T_K$ that scales with $\Delta$ alone. When the grain is so large
that the localization length $\xi$ is smaller than $L$, $E_c$ and
$\Delta$ are irrelevant and $T_K$ is determined by the average energy
level spacing in the vicinity of the magnetic impurity, $\Delta_\xi =
1/\nu\, \xi^d$ \cite{alk95}. Thus, one expects RMT to be applicable on
the scale $\Delta_\xi$ in this local random matrix theory (LRMT)
regime. However, corrections to RMT and LRMT due to correlations of
wave functions of order $1/g$ are found to be essential for
determining $T_K$. Finally, when $T_K$ is smaller than $\Delta$ or
$\Delta_\xi$, the distribution of $T_K$ is mainly determined by the
coupling of the magnetic impurity to a two-level system.


In the symmetric Anderson model of a magnetic impurity \cite{anderson}, the
exchange interaction matrix elements are estimated as $J_{kl} =
4 t_{ik}^\ast t_{il}/U$, where $U$ is the on-site Coulomb energy and
$t_{ik}$ the hopping matrix elements connecting the localized state
$\phi_i({\bf r})$ with a delocalized state $\psi_k({\bf r})$: $ t_{ik}
= \int d^dr\, \phi_i^\ast({\bf r})\, \hat{T}\, \psi_k({\bf r})$, where
$\hat{T}$ is the kinetic energy operator. For states localized at
${\bf r}=0$ with a radius $a_0$, $ J_{kl} \approx J\, \psi^\ast_k(0)\,
\psi_l(0), $ where $J= 4 (1/m^\ast a_0^2)^2/ U$ is the exchange
coupling, $m^\ast$ the band mass. In the antiferromagnetic case
($J<0$), $J$ is strongly renormalized. The temperature at which the
second-order correction is equal to the bare value defines $T_K$
\cite{nagaoka},
\begin{equation}   
\label{eq:tk}  
1 = \frac{1}{2x} \sum_{n=1}^N \frac{x_n}{s_n} \tanh \left(  
\frac{s_n}{2 \kappa} \right),
\end{equation}   
where $N$ is the number of states in the band of width $D$. Here, we
have used the rescalings $x=D/J$ and $\kappa = T_{\rm K}/\Delta$, with $x_n
= L^d |\psi_n ({\bf r}) |^2$ equals the probability density of the
$n$-th eigenstate and $s_n = (E_n-E_F)/\Delta$ is the eigenenergy
relative to the Fermi energy in units of $\Delta$. The number of
electrons in the Kondo box is taken to be even.
Equation (\ref{eq:tk}) is valid in the weak coupling regime, when $T_K
\ll D$. It coincides with the self-consistent solution of the one-loop
renormalization group (RG) equation \cite{anderson,wilson} up to the
$\tanh$ factor, which accounts for the finite-temperature occupation.
While the approximations involved in deriving Eq. (\ref{eq:tk}) fail
to describe the Kondo system below $T_K$, Eq. (\ref{eq:tk}) does yield
the correct low-energy scale in the effective Fermi liquid theory,
where $T_K$ determines the Landau parameters
\cite{nozieres}. Nevertheless, for finite systems, there may be
deviations from one-parameter scaling due to sample-to-sample
mesoscopic fluctuations \cite{kbu041}.
%
%

In order to gain insight on the statistical properties of $T_K$ over a
wide range of values of $J$, we solved Eq. (\ref{eq:tk}) numerically
and compared the resulting distribution and moments of $T_K$ with
approximate analytical expressions appropriate to the different
regimes already mentioned.

\begin{figure}[ht]  
\includegraphics[width=7cm]{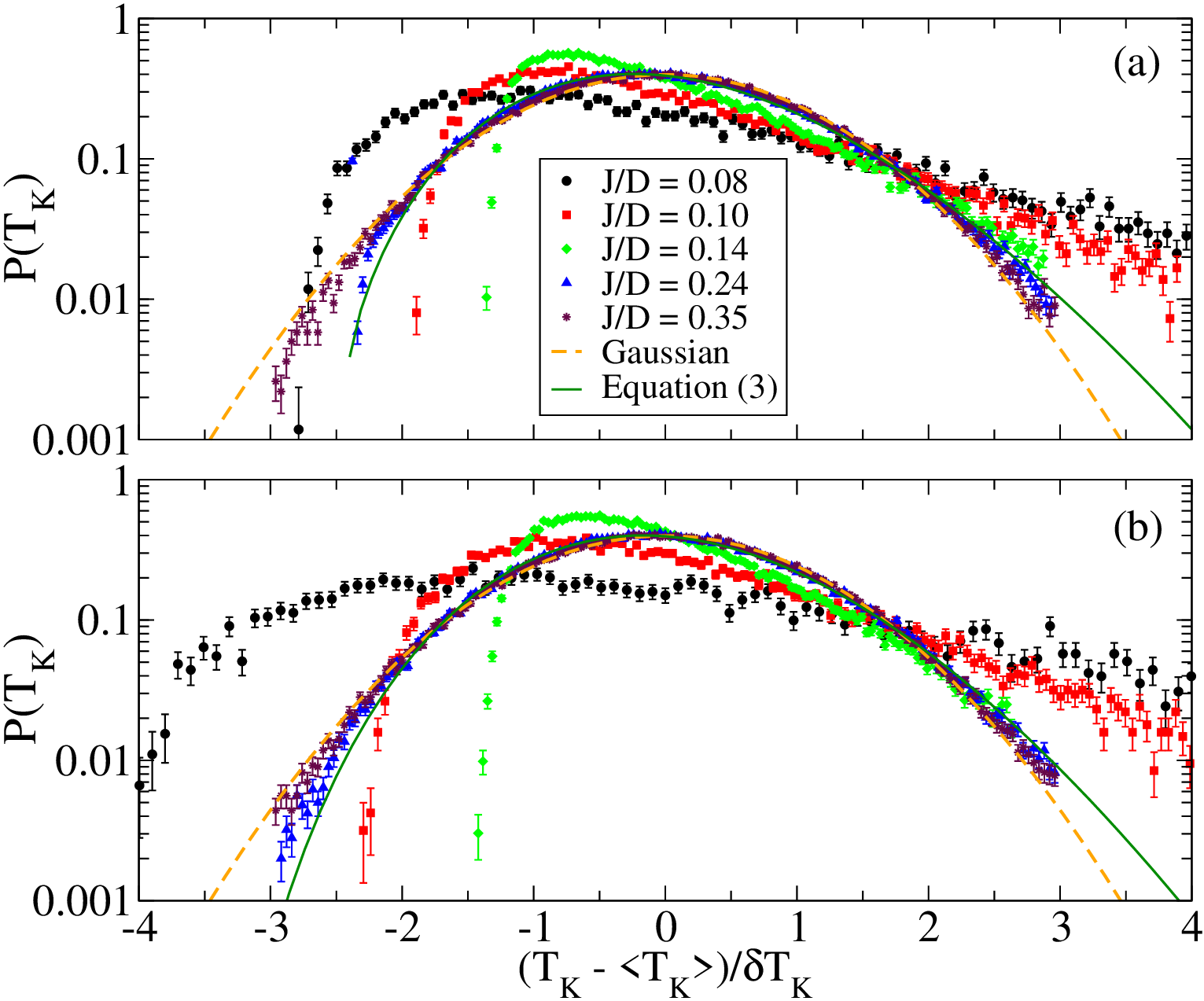}  
\caption{(Color on-line) The distribution of $T_K$ as scaled with its
standard deviation $\delta T_K $ for GOE (a) and GUE (b). Events
corresponding to $T_K=0$ were neither included in the distributions
nor in the average and the standard deviation used in the shifting and
rescaling of $T_K$. The dashed line corresponds to a Gaussian with
zero mean and unit variance. The solid line corresponds to
Eq. (\ref{eq:PTKlarge}) for $J/D=0.24$.}
\label{fig:TK_dis}  
\end{figure}  

{\it Random Matrix Theory.} For $g \gg 1$, the energy levels within an
energy window smaller than $E_c$ obey the Wigner-Dyson distribution of
RMT \cite{efetov}. The eigenenergies $s_n$ and eigenfunction
intensities $x_n$ fluctuate independently. The latter are themselves
uncorrelated and obey the Porter-Thomas distributions
\cite{efetov,mirlin}. Figure \ref{fig:TK_dis} shows the distribution
of $T_K$ obtained from the unfolded spectrum of 
random matrices of size $N=500$ using 500
realizations for each Gaussian ensemble, namely orthogonal (GOE) and
unitary (GUE). The Fermi energy is taken to be in the middle of the band.
The dependence of the average $\langle T_K \rangle$ on $J$ at $\langle
T_K \rangle > \Delta $ is found to coincide with the clean case
result, which is given by $T_K^{(0)} \approx 0.57\, D \exp ( - D/J)$
for a spectrum of equally spaced levels. For small $J$, $T_K$
approaches $\Delta$; in a  system without fluctuations of energy levels and wavefunctions, it would turn abruptly to zero
at $ J^\ast \approx D/[\ln (2 N) + 0.58]$ in a nonanalytical fashion,
$T_K^{(0)}(J\rightarrow J^\ast) = -\Delta/\ln \left[(D/J^\ast -
D/J)/4\right]/2$. However, fluctuations are important at $\langle T_K
\rangle < \Delta$, leading to non-zero values for $T_K$ below the
clean-limit threshold $J^\ast$. In Fig. \ref{fig:TK_dis}, the
distributions of Kondo temperatures for the GOE and GUE are shown for
different values of $J$. The Kondo temperatures are shifted and
rescaled in order to facilitate the comparison with a Gaussian
distribution. The latter might be expected from a naive use of the
central limit theorem: Since wave function amplitudes corresponding to
distinct eigenstates are uncorrelated in RMT, $T_K$ is determined by a
sum of independent random variables when energy level fluctuations are
neglected. However, even when $\langle T_K \rangle$ exceeds $\Delta$
there are deviations from the Gaussian behavior. This is in agreement
with approximate analytical results obtained by performing the average
over the wave function amplitudes exactly but assuming equally spaced
energy levels and $T_K \gg \Delta$, when we find in the limit of large number of levels $N$ in a straight forward but lengthy derivation\cite{longpaper}
\begin{equation}  
\label{eq:PTKlarge}  
P \left( T_K \right) \sim \frac{1}{\Delta}\exp \left[  
\beta\kappa \ln \left( \frac{e \kappa_0}{\kappa} \right)  
-\beta\kappa_0 \right],  
\end{equation}  
where $\kappa_0 = T_K^{(0)}/\Delta$ and $\beta$ denotes the ensemble
($\beta=1$ for the GOE and 2 for the GUE). This distribution yields
$\langle T_K \rangle =T_K^{(0)}$, independent of $\beta$. For $T_K
\approx T_K^{(0)}$, it is approximately Gaussian with a width $ \delta
T_{K\, \beta} = \sqrt{\Delta\, T_K^{(0)}/\beta}$, in agreement 
 with an earlier derivation \cite{moriond}, but note that 
  the distribution is not log-normal as was incorrectly concluded, there. The ratio between
the variance in the orthogonal and unitary regime is thus $\sqrt{2}$
for $T_K^{(0)} > \Delta$, in agreement with numerical results
\cite{longpaper}. When time-reversal symmetry is present, the energy
level repulsion is weaker and the tendency to localization stronger,
resulting in a wider distribution than in the unitary regime
\cite{lewenkopf05}. Within RMT, the width is found to vanish in the
thermodynamic limit, when $\Delta \rightarrow 0$\cite{moriond}.


\begin{figure}[ht]  
\includegraphics[width=6.25cm]{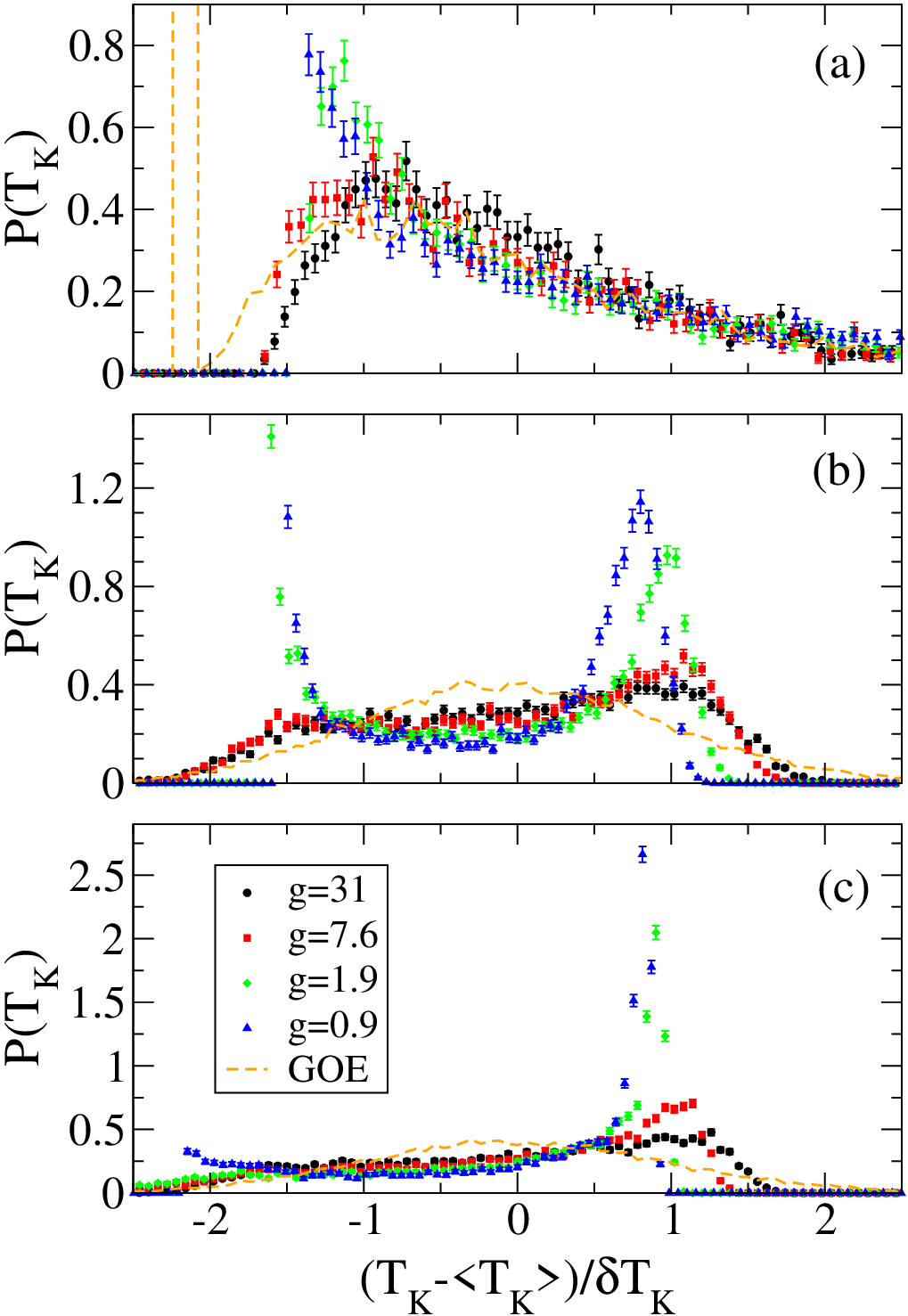}  
\caption{(Color on-line)    The distribution of $T_K$ with eigenstates of
a $20\times 20$ Anderson model without magnetic field (solid symbols)
compared with the distribution for the GOE with $N=400$ (dashed
line). $J/D =$ (a) 0.1, (b) 0.35, and (c) 0.6. For the Anderson model,
4,000 realizations were used for each value of $g$.}
\label{fig:TK_dist}  
\end{figure}  

\vspace{-.15cm}
  
{\it Anderson model.} Going beyond RMT, stronger deviations from the
Gaussian behavior can be expected due to correlations in the wave
functions. In order to describe the conduction electrons, we use the
noninteracting tight-binding Anderson model with nearest-neighbor
hopping $t$ and random site energies drawn independently from a box
distribution of width $W$ centered at zero. We assume each
single-electron eigenstate to be spin degenerate and consider
$20\times 20$ square lattices with periodic boundary conditions. The
Hamiltonian was diagonalized and the eigenenergies and eigenvectors
were used to determine $T_K$ through Eq. (\ref{eq:tk}) without
unfolding of energy levels. The disorder amplitude $W$ and the
parameter $g$ are related in the Born approximation by $1/g =
\frac{\pi}{48} W^2/(E_F t)$, where the lattice constant is $a=1$
($N=L^2$) and $\Delta = 8t/N$. This expression is valid for weak
disorder $g>1$ and for a flat density of states. We chose $E_F =2t$ to
avoid traces of the van Hove peak at the band middle. Only
time-reversal symmetric cases where studied. The resulting
distribution of $T_K$ is shown in Fig. \ref{fig:TK_dist}, where a
non-Gaussian shape is evident. The GOE and Anderson model
distributions only agree at weak disorder and for $J \ll D$,
Fig. \ref{fig:TK_dist}(a). For larger $J$ [see
Fig. \ref{fig:TK_dist}(b)], the distribution remains non-Gaussian for
all disorder strengths, while the GOE distribution corresponding to
the same value of $J/D$ is almost Gaussian. The disordered Anderson
model presents a bimodal structure in $P(T_K)$ that is absent in the
RMT case. That structure is strongly pronounced in the strong disorder
regimes, with two distinct peaks clearly visible. As $J/D$ increases
further, weight is transfered from the low-$T_K$ to the high-$T_K$
peak [see Fig. \ref{fig:TK_dist}(c)].

Analytical information on the fluctuations of $T_K$ can be obtained by
relating it to the correlation function of the LDOS
\cite{mirlin}. Defining $T_K^{(0)}$ by Eq. (\ref{eq:tk}) with average
DOS $\nu$ , we can rewrite Eq. (\ref{eq:tk}) in terms of the LDOS as
\cite{zu96}
\begin{equation}  
\label{eq:TK_transcd}  
\ln \frac{T_K}{ T_K^{(0)}}= \int_{-E_F}^{D-E_F} dE\, \frac{\delta  
\rho(E_F + E, {\bf r})}{2 \nu E} \tanh \left( \frac{E}{2 T_K} \right).  
\end{equation}  
Expanding $T_K$ to second order in $\delta \rho = \rho - \nu$, we can
relate it to the correlation function of LDOS, $R_2^{\rm LDOS}(\omega)
= \langle \rho({\bf r},E)\, \rho ({\bf r}, E + \omega)
\rangle/\nu^2$. While in RMT there are no correlations between wave
functions at different energies, in the disordered Anderson model they
are of order $1/g$,
\begin{equation}  
\label{psilk}  
L^2 \left\langle |\psi_{n}({\bf r})|^2 |\psi_{m} ({\bf r})|^2
\right\rangle \equiv f_{nm}(\omega) = 1 + \frac{2}{\beta} {\rm Re}\,
\Pi_\omega,
\end{equation}  
independently of the states $n,m$ ($n\neq m$). The correlation is
stronger in the time-reversal symmetry case. The dependence on
disorder is determined by a sum over diffuson modes, yielding, for
$L>l$ and in two-dimensions (2D),
\begin{equation}  
{\rm Re}\, \Pi_\omega = \frac{1}{4 \pi g} \ln \left( \frac{1/4\tau^2 +  
\omega^2}{ E_c^2 + \omega^2} \right).  
\end{equation}  
In terms of the spectral correlation function, $R_2(\omega)=\langle
\rho(E)\rho(E+\omega)\rangle/\nu^2-1$, the correlation function of
LDOS reads
\begin{equation}  
R_2^{\rm LDOS}(\omega) = R_2 (\omega)\, f_{nm}(\omega) +
\delta(\omega/\Delta)\, f_{nn}(\omega).
\end{equation}  
For $\omega<E_c$, $R_2 (\omega)$ has an oscillatory correction of
order $1/g^2$ to the leading RMT term \cite{mirlin}. For $\omega>E_c$,
the oscillatory part of the spectral correlation function decays
exponentially, while there is a correction of order $1/g^2$ which
decays like $1/\omega$ without oscillations \cite{mirlin}. Expanding
the exponent to second order in $\delta \rho$, we find
%
%
\begin{equation}
\label{eq:dTK_g}  
(\delta T_K)^2 = 2 \left( T_K^{(0)} \right)^2 \left[ S_{\beta} \left(
\tau,T_K^{(0)} \right) + C_{\beta} \left( \tau, T_K^{(0)} \right)
\right],
\end{equation}
where $S_{\beta}$ arises from the spectral self-correlation term
proportional to $\delta(\omega/\Delta)$ and is of order $\Delta/T_K$.
To this order we also obtain $\langle T_K \rangle = T_K^{(0)} +(\delta
T_K)^2/2 T_K^{(0)}$. The decaying parts of the spectral correlation
function and the correlations of wave functions yield
\begin{eqnarray}  
\label{eq:Ccorr}  
C_{\beta} (\tau, T_K) & = & \int_{-E_F}^{D-E_F} \frac{dE\,  
dE^\prime}{8\, E\, E^\prime}\, \tanh \frac{E}{2 T_K} \, \tanh  
\frac{E^\prime}{2 T_K} \nonumber \\ & & \times \left[ -1 +  
R_2(E-E^\prime) f_{nm}(E-E^\prime)\right].  
\end{eqnarray}  
In the thermodynamic limit $E_c \rightarrow 0$, the leading term in
Eq. (\ref{eq:Ccorr}) arises due to the correlations between wave
functions; to leading logarithmic order, we obtain
\begin{equation}  
\label{eq:rbeta}  
C_{\beta} (\tau, T_K) = \frac{1}{6 \pi \beta g} \ln^3 \left(
\frac{E_F}{g T_K} \right),
\end{equation}  
which is valid for $D/J \gg \ln g$. Thus, the width of the
distribution is finite in the thermodynamic limit and the deviation of
the average Kondo temperature from the clean limit is stronger than
expected from Anderson's theorem when applied to the Kondo problem, in
contrast to an earlier prediction \cite{chakravarty00}. A finite
$\delta T_K$ will have implications to the thermodynamic properties of
disordered metals with magnetic impurities, as well as to heavy
fermion materials. An earlier work \cite{kotliar} considered the
effect of fluctuations of LDOS as well, but obtained an even larger
effect by disregarding that wave function correlations are only of
order $1/g$.

\begin{figure}[ht]  
\includegraphics[width=6cm]{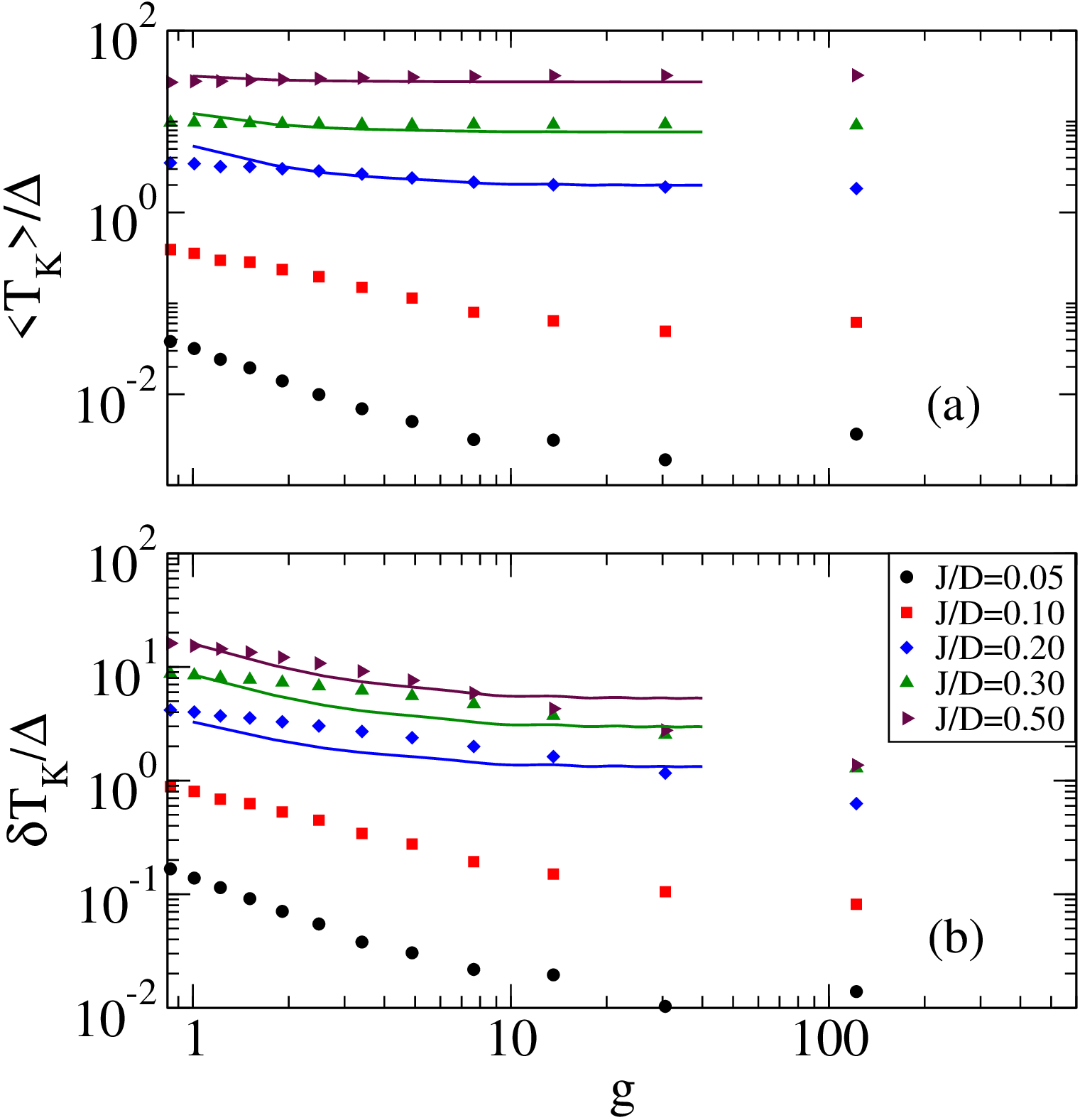}  
\caption{(Color on-line) Dependence of average Kondo temperature
$\langle T_K \rangle$ (a) and standard deviation (b) in the Anderson
model on the dimensionless conductance $g$ for different values of
$J/D$. Conditions are identical to Fig. \ref{fig:TK_dist}, with
averages taken over 500 realizations. The solid lines correspond to
Eq. (\ref{eq:dTK_g}) keeping $E_c$ and $\Delta$ finite. Notice that
since we assume $E_c\tau<1$, the lines are only presented for
$1<g<20$.}
\label{fig:dTK0}  
\end{figure}  

\vspace{-.15cm}

The expressions for $\langle T_K \rangle$ and $\delta T_K$ based on
Eq. (\ref{eq:dTK_g}) are compared with results from the numerical
simulations in Fig. \ref{fig:dTK0}. They are found to agree
qualitatively with the dependence on the disorder strength $g$ seen in
the simulations for different values of $J$. It is only at $g \gg 1$
and $\langle T_K \rangle \gg \Delta$ that $\langle T_K \rangle$ is
insensitive to disorder. In general, for weak disorder, there is
good agreement between GOE and Anderson model results for $\delta T_K$
(not shown). At stronger disorder, fluctuations in the Kondo
temperature increase in amplitude due to the increase of correlations
in the LDOS at different energies over an interval of order of
$1/\tau$, Eq. (\ref{psilk}).

\begin{figure}[ht]  
\includegraphics[width=6.5cm]{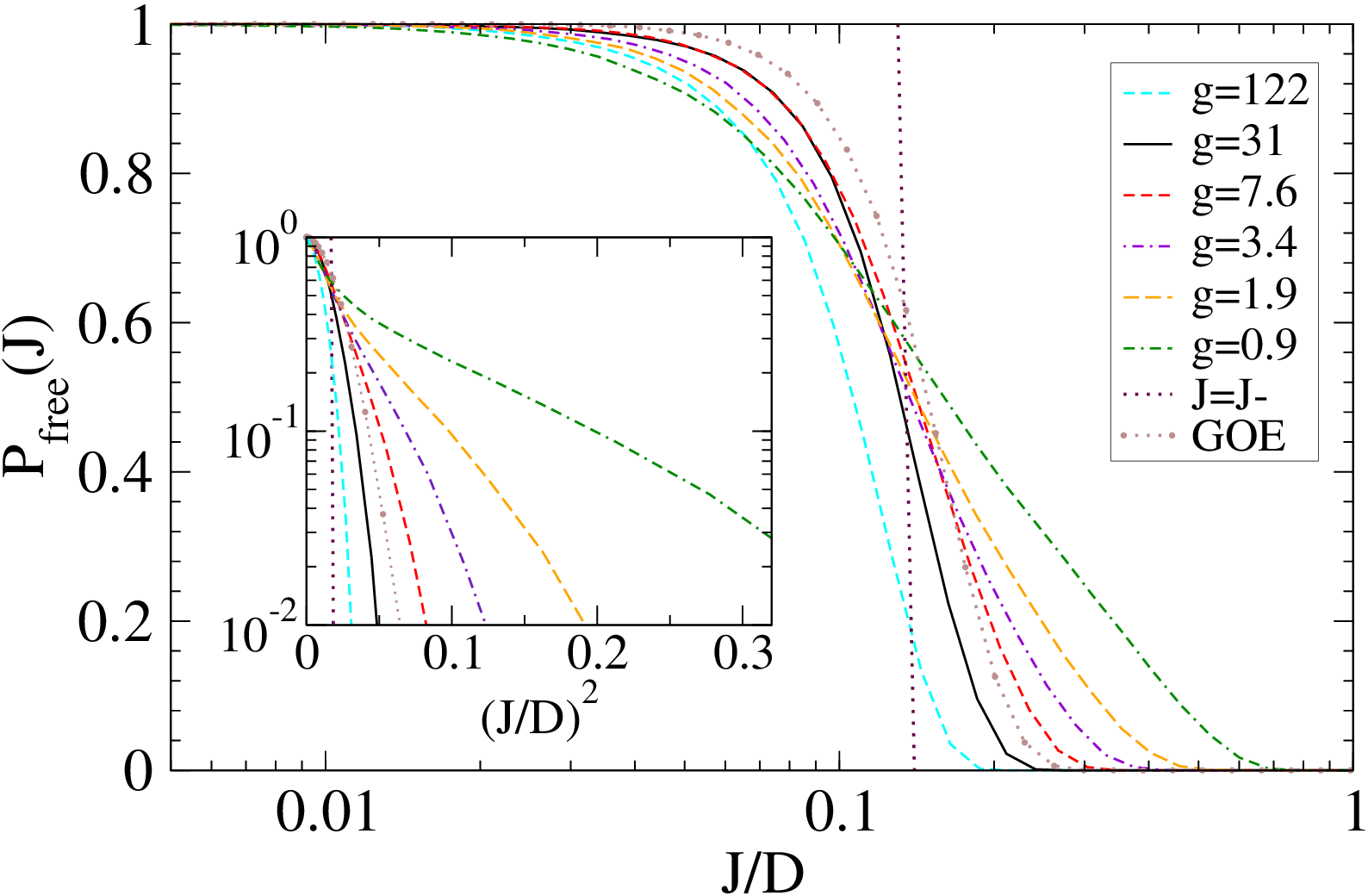}  
\caption{(Color on-line) Probability of free moments as a function of  
$J$ for the Anderson model at different disorder strengths $g$ (same  
conditions of Fig. \ref{fig:dTK0}). Also shown are the numerical  
result for the GOE with $N=400$. $J^\ast$ was estimated for a flat band.}  
\label{fig:TK0}  
\end{figure}  


{\it Free moments.} The long tail toward small-$T_K$ indicates that
disorder enhances the probability of having free magnetic moments at
low temperature. This is confirmed in Fig. \ref{fig:TK0}, where we
show the probability of finding no solution to Eq. (\ref{eq:tk}) as a
function of $J/D$. We note that for $T_K < \Delta$ only the two
closest levels to the Fermi energy are strongly coupled to the
magnetic impurity.
In the clean limit, the probability of free moments is a sharp step
function of J, namely, $P_{\rm free}(J \le J^\ast)= 1$, and $P_{\rm
free}(J > J^\ast)=0$. When disorder is present, $J^\ast$ fluctuates 
due to the random wave function intensities $x_n$ and level spacings
$s_n$ of the two closest levels to the Fermi energy
\cite{longpaper}. The probability that the level spacing at the Fermi
energy is of the order of $T_K^{(0)} (J)$ is exponentially small for
$J > J^\ast$. However, in Fig. \ref{fig:TK0} we see that the decay
with $J$ is slower than that, indicating that the fluctuations of wave
functions are crucial in the appreciable increase in the probability
of free moments for $J > J^\ast $.


In the absence of an external magnetic field, magnetic impurities can
break time-reversal symmetry since the spin dynamics of the magnetic
impurities is slow compared to the time scale of the conduction
electrons \cite{hikami}. For a mesoscopic sample at $T>E_c$, weak
localization corrections are suppressed when $ X_s^{\rm WL} =
1/E_c\tau_s$ exceeds $1$ \cite{hikami}. At finite temperature, the
spin scattering rate $1/\tau_s$ is renormalized by the Kondo
correlations: It is small at both $T\gg T_K$ and $T\ll T_K$, having a
maximum at around $T=T_K$ \cite{bergmann}, which is given by a
fraction $\alpha$ of the unitary limit of the scattering cross
section. In 2D, one finds $1/\tau_{s \rm max} = \alpha\, n_M v_F/(2
k_F)$, where $n_M$ is the concentration of magnetic impurities, with
$\alpha \approx 0.2$ \cite{zarand,baeuerle}. Thus, the maximum value of the
crossover parameter is given by $ X_{s M} = \alpha N_M/g$, where $N_M
= n_M L^2$ is the number of magnetic impurities in the sample. When
there are only a few magnetic impurities, $N_M < g$, $X_{s} < 1$, the
metallic grain is in the orthogonal regime. Increasing the
concentration of magnetic impurities increases the parameter $X_s$,
thus decreasing the width of the distribution of Kondo
temperatures. The maximum in the temperature dependence of the spin
scattering rate results in a plateau of the dephasing time
\cite{bergmann}.
Mesoscopic Au wires with Fe impurities at concentrations $n_S = 3-10$
ppm have a Kondo temperature $T_K \approx 300$ mK \cite{bergmann}.
While both $E_c$ and $\Delta$ are found to be of order $10^{-7}$K, and
thus completely negligible in such wires, we can estimate that $J/D =
0.074$ is much larger than $J^\ast/D = 0.035$. Therefore, we obtain
from Eq. (\ref{eq:rbeta}) that the width of the distribution of Kondo
temperatures $\delta T_K$ is of the same order as $\langle T_K
\rangle$, with $ \langle T_K \rangle= 300$ mK.

To conclude, the distribution of Kondo temperatures has a finite width
in disordered metals due the correlations of eigenfunctions within an
energy interval of width $1/\tau$. Thus, the presence of free magnetic
moments in disordered metals at temperatures below the average Kondo
temperature is expected to yield a finite contribution to the electron
dephasing rate and to the thermodynamic properties of disordered
metals in the presence of localized magnetic moments \cite{miranda}.
  
  
The authors gratefully acknowledge useful discussions with
H. Baranger, C. Beenakker, C. Chamon, P. Fulde, R. Kaul, C. Lewenkopf,
E. Miranda, G. Murthy, M. Raikh, and D. Ullmo. The authors thank the
hospitality of the Condensed Matter Theory Group at Boston University.
This research was supported by the German Research Council (DFG) under
SFB 508, A9, and by the EU TMR-network, Grant HPRN-CT2000-0144.
Work at UCF was supported in part by NSF CCF-0523603 and
 by the Interdisciplinary Information Technology Laboratory
 (I2Lab).

{\it Note:} After the submission of this work we learned that a result
similar to Eq. (\ref{eq:rbeta}) was recently obtained independently by
Micklitz and collaborators \cite{micklitz05}.
After the submission we  learned of the work by Cornaglia and
collaborators who also studied numerically the distribution of the
  Kondo temperature with the same approach, and  found a peak at small $T_K$, as well \cite{grempel}.   


  
\end{document}